\documentclass[10pt,letterpaper]{article}
\usepackage{opex3}
\usepackage{cite}
\usepackage{amsmath}
\usepackage{amssymb}
\usepackage{setspace}

\begin{document}

\title{Stimulated brillouin scattering in slow light waveguides}
\author{Wenjun Qiu,$^1$ Peter T. Rakich,$^{2,3}$ Marin Solja\v{c}i\'{c},$^1$ and Zheng Wang$^{4*}$}
\address{
$^1$Department of Physics, Massachusetts Institute of Technology, Cambridge, MA 02139 USA \\
$^2$Sandia National Laboratories, PO Box 5800 Albuquerque, NM 87185 USA\\
$^3$Department of Applied Physics, Yale University, New Haven, CT 06520 USA\\
$^4$Department of Electrical and Computer Engineering, University of Texas at Austin, Austin, TX 78758 USA}
\email{$^*$zheng.wang@austin.utexas.edu}

\begin{abstract}
We develop a general method of calculating Stimulated Brillouin Scattering (SBS) gain coefficient in axially periodic waveguides. Applying this method to a silicon periodic waveguide suspended in air, we demonstrate that SBS nonlinearity can be dramatically enhanced at the brillouin zone boundary where the decreased group velocity of light magnifies photon-phonon interaction. In addition, we show that the symmetry plane perpendicular to the propagation axis plays an important role in both forward and backward SBS processes. In forward SBS, only elastic modes which are even about this plane are excitable. In backward SBS, the SBS gain coefficients of elastic modes approach to either infinity or constants, depending on their symmetry about this plane at $q=0$.
\end{abstract}


\bibliographystyle{osajnl}
\bibliography{phononics}

\section{Introduction}

Stimulated Brillouin Scattering (SBS) is a third order nonlinear process in which two optical modes are coupled through an elastic mode \cite{Boyd,Agrawal}. In a waveguide system, the interference of pump and Stokes waves generates dynamic optical forces at the beat frequency. The optical force, while on resonance with an elastic mode at the phase-matching wavevector, excites the mechanical vibration in the waveguide, which in turn scatters light between the pump and Stokes waves. Since its discovery, SBS has been extensively studied with a variety of applications in efficient phonon generation \cite{Chiao64,Russell06}, optical frequency conversion \cite{Kobyakov10,Russell09,Pant11}, slow light \cite{Song05,Song06,Gaeta05,Pant08} and signal processing techniques \cite{Carmon09,Gauthier07}.

Previous experimental and theoretical studies of SBS have heavily focused on axially invariant systems, such as optical fibers, phononic crystal fibers, and rectangular waveguides \cite{Kobyakov10,Gauthier11,Russell06,Russell09,Russell10,Russell11,Pant11,PRX}. It is desirable to study the SBS process in axially periodic waveguide for two reasons. First, the unit cell in periodic structures provides more degree of freedoms to tailor the distributions of optical force and elastic deformation \cite{PRX}. More importantly, the slow group velocities of light, which are readily achievable in periodic waveguides, can dramatically enhance SBS nonlinearities \cite{Soljacic04}. In axially invariant waveguides, all the vectorial fields can be decomposed into real/imaginary-valued longitudinal and transverse components. Such decomposition are useful in characterizing various SBS processes \cite{PRX}. In axially periodic waveguides, all the components of electric field and elastic deformations become complex-valued. A careful analysis of the symmetry in optical forces and elastic modes is required to elucidate photon-phonon coupling and the resultant selection rules in various SBS processes.

In this article, we propose a general method of calculating SBS gains in periodic waveguides. Armed with this general formalism, we study the forward SBS (FSBS) and backward SBS (BSBS) processes of a suspended silicon periodic waveguide. We demonstrate that the strength of both FSBS and BSBS can be dramatically enhanced in the slow light regime. In addition, we show that the symmetry plane perpendicular to the propagation axis plays an important role in characterizing different SBS processes. This plane separates elastic modes at $q=0$ into even and odd modes. For FSBS, only even modes are excitable. For BSBS, the gain coefficients approaches to either infinity or constants, depending on the mode symmetries at $q=0$.

\section{Calculating the SBS gain of periodic waveguides}

To start with, we develop a general method of calculating the SBS gain of periodic waveguide systems. The optical (elastic) eigen-modes of a waveguide with axial periodicity $a$ is characterized by Bloch wavevector $\mathbf{k}$ ($\mathbf{q}$). In a typical SBS process, the pump wave $\mathbf{E}_p e^{-i\omega_p t}$ at wavevector $\mathbf{k}_p$ and Stokes wave $\mathbf{E}_s e^{-i\omega_s t}$ at wavevector $\mathbf{k}_s$ generate optical forces varying in space at wavevector $\mathbf{q}=\mathbf{k}_p - \mathbf{k}_s$ and oscillating in time at the beat frequency $\Omega = \omega_p - \omega_s$. This optical force can excite mechanical vibrations which enables the parametric conversion between pump and Stokes waves. Assuming the propagation direction is parallel to the $x$ axis, this process can be described by the following relation \cite{Boyd}
\begin{equation}
\frac{dP_s}{dx} = gP_pP_s - \alpha_s P_s
\end{equation}
Here, $P_p$ and $P_s$ are the guided power of the pump and Stokes waves, and $g$ is the SBS gain. Through particle flux conservation, SBS gain is given by the follow formula \cite{PRX}:
\begin{equation}
g(\Omega) = \frac{\omega_s}{2\Omega a P_p P_s} Re\left\langle \mathbf{f}, \frac{d\mathbf{u}}{dt} \right\rangle 
\end{equation}
where $\mathbf{f}$ is the optical force generated by pump and Stokes waves, and $\mathbf{u}$ is the elastic response of the waveguide induced by $\mathbf{f}$. The optical power of a periodic waveguide is given by $P=v_g \langle \mathbf{E}, \epsilon \mathbf{E} \rangle /2a$, where $v_g$ is the optical group velocity. Therefore,
\begin{equation}
g(\Omega) = \frac{2\omega_s a}{v_{gp}v_{gs}}
\frac{Im \langle \mathbf{f},\mathbf{u} \rangle}{\langle \mathbf{E_p}, \epsilon \mathbf{E_p} \rangle \langle \mathbf{E_s}, \epsilon \mathbf{E_s} \rangle}
\end{equation}
The elastic response $\mathbf{u}$ can be decomposed into elastic eigen-modes at $\mathbf{q}$, $\mathbf{u}=\sum_m b_m \mathbf{u}_m$ with $b_m$ given by Eq. (\ref{b_m}). The total SBS gain is the sum of SBS gains of individual elastic modes:
\begin{equation}
g(\Omega) = \sum_m G_m \frac{(\Gamma_m/2)^2}{(\Omega - \Omega_m)^2 + (\Gamma_m/2)^2}
\end{equation}
The SBS gain of the $m^{th}$ elastic mode has a Lorentian shape and a peak value $G_m$:
\begin{equation}
G_m = \frac{2 \omega_s a}{\Omega_m \Gamma_m v_{gp} v_{gs}}
\frac{|\langle \mathbf{f},\mathbf{u}_m \rangle|^2}
{\langle \mathbf{u}_m \rho \mathbf{u}_m \rangle \langle \mathbf{E_p}, \epsilon \mathbf{E_p} \rangle \langle \mathbf{E_s}, \epsilon \mathbf{E_s} \rangle}
\label{G_m}
\end{equation}
In Eq. (\ref{G_m}), the overlap integrals of different optical forces are linearly summed. Here, we consider three kinds of optical forces: electrostriction body force, electrostriction pressure, and radiation pressure \cite{Wang10,PRX}. Electrostriction body force is integrated over the volume of the unit cell, while electrostriction pressure and radiation pressure are integrated over the boundary of the unit cell. Equation (\ref{G_m}) shows that the SBS gain coefficient is dependent on the frequency ratio, the mechanical loss factor, the optical group velocities of pump and Stokes waves, and the overlap integral between optical forces and elastic eigen-modes. With periodic waveguides with unit cell structures, not only can the overlap integral be tailored, we can also take advantage of the decreased optical group velocity near the brillouin zone boundary to gain further enhancement of SBS nonlinearities.

\section{Optical and elastic modes of a silicon periodic waveguide}

\begin{figure}[p]
\centering\includegraphics[width=\textwidth]{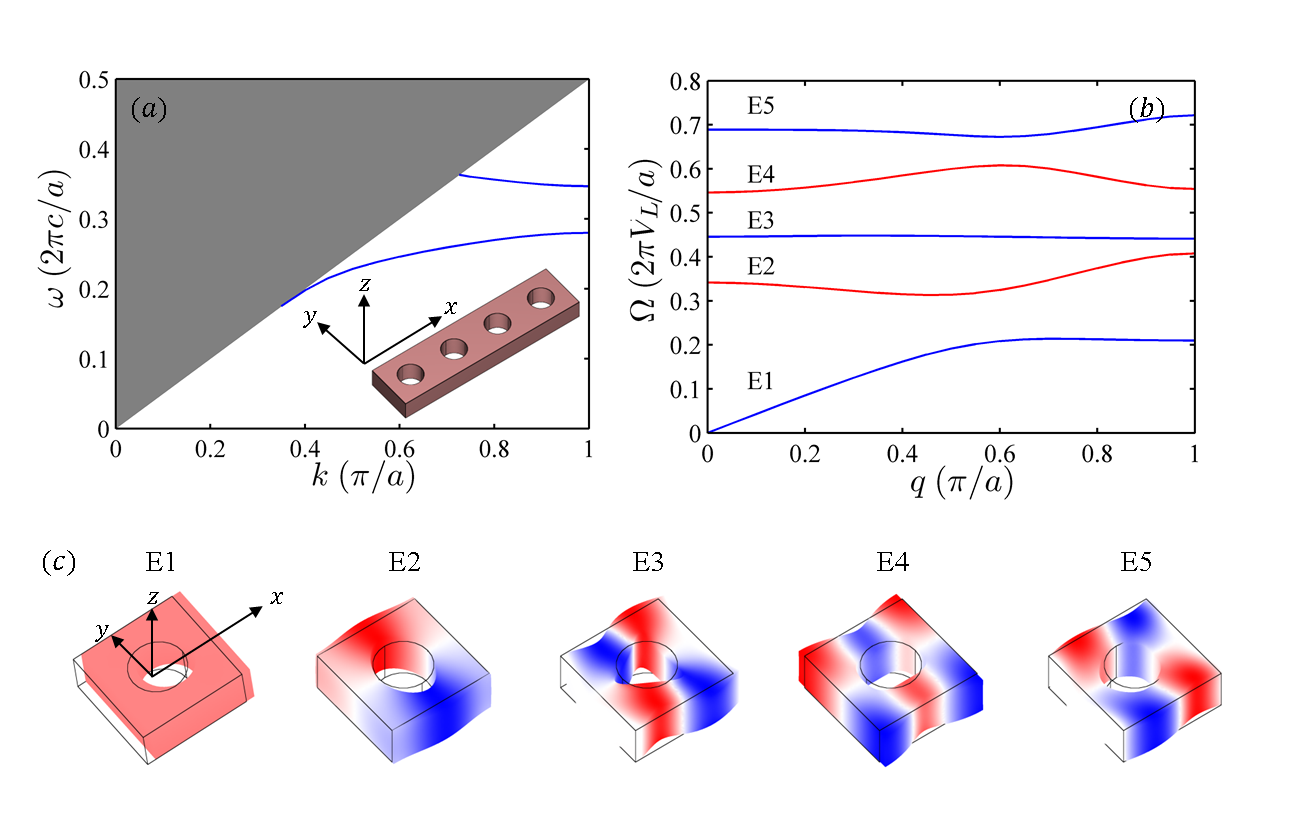}
\caption{The optical and elastic modes of a silicon periodic waveguide.  (a) The dispersion relation of the fundamental optical mode which is yodd and zeven. The area shaded in gray represents light cone of air. (b) The dispersion relations of elastic modes which are yeven and zeven. Xeven and xodd modes at $q=0$ are colored in red and blue respectively. (c) Elastic modal profiles at $q=0$. The deformation is proportional to the displacement $\mathbf{u}$. The colored surface represents displacement component $u_y$. Red, white and blue correspond to positive, zero, and negative values respectively. Mode E1 experiences a parallel shift along $x$ direction.}
\label{fig1}
\end{figure}

Using the formalism developed above, we proceed to study the SBS process of a a suspended silicon waveguide with periodic cylindrical holes (Fig. \ref{fig1}(a) insert). The axial periodicity is $a$, the cross-section in $yz$ plane is $a$ by $0.4a$, and the radius of the cylindrical air hole is $0.25a$. For silicon, we use refractive index $n=3.5$, Young's modulus $E=170\times 10^9$ Pa, Poisson's ratio $\nu=0.28$, and density $\rho=2329$kg/m\textsuperscript{2}. In addition, we assume that the [100], [010], and [001] symmetry direction of this crystalline silicon coincide with the $x$, $y$, and $z$ axis respectively. Under this orientation, the photo-elastic tensor $p_{ijkl}$ in the contracted notation is $[p_{11},p_{12},p_{44}] = [-0.09,0.017,-0.051]$ \cite{Briddon06}. The waveguide has three symmetry planes $x=0$, $y=0$, and $z=0$. We assume that the crystalline structure of the material is also symmetric about these three planes so that the anisotropy in optical, elastic, and photo-elastic constants doesn't break these mirror symmetries. This condition is clearly satisfied for silicon in the current orientation.

First, we analyze the optical modes of the waveguide. The optical modes are categorized into yeven/yodd (zeven/zodd) according to their symmetries about plane $y=0$ ($z=0$). The fundamental mode is yodd and zeven with $E_y$ as the dominant component of electric field (Fig. \ref{fig1}(a)). We fix the pump wavelength at $1.55\mu$m. So a different operating point in the dispersion relation indicates a different value of $a$. The optical mode $\mathbf{E}$ doesn't have symmetry about plane $x=0$ since nonzero $k$ breaks this mirror symmetry. Actually, the mirror reflection of eigen-mode $\mathbf{E}$ at $k$ corresponds the eigen-mode at $-k$, which is also the complex conjugate of $\mathbf{E}$. Under a properly chosen phase, the mirror reflection of $\mathbf{E}$ is exactly $\mathbf{E}^*$. We can write this relation as:
\begin{equation}
E_i(-x,y,z) = E_i^*(x,y,z)s_i
\label{PE}
\end{equation}
where $s_x=-1$, and $s_{y,z}=1$. The subscript in $s$ does not introduce summation when encountered with repeated indices.

Next, we analyze the elastic modes of the waveguide. Again, the elastic modes are categorized into yeven/yodd (zeven/zodd) based on their symmetries about $y=0$ ($z=0$). In intra-modal coupling where both pump and Stokes waves reside in the same optical eigen-mode, the optical force is always symmetric with respect to planes $y=0$ and $z=0$. Therefore, we only need to consider E-modes which are both yzeven and zeven (Fig. \ref{fig1}(b)). At $q=0$, the symmetry about plane $x=0$ is recovered, separating E-modes into xeven and xodd modes (Fig. \ref{fig1}(c)). In addition, the elastic eigen-equation at $q=0$ is invariant under conjugation operation, resulting in real-valued $\mathbf{u}$. At nonzero $q$, the elastic modes are neither symmetric nor anti-symmetric about plane $x=0$, and $\mathbf{u}$ are complex-valued. Similar to the optical modes, we can choose a proper phase of $\mathbf{u}$ so that the mirror reflection of $\mathbf{u}$ is exactly $\mathbf{u}$:
\begin{equation}
u_i(-x,y,z) = u_i^*(x,y,z)s_i
\label{Pu}
\end{equation}

\section{Forward SBS}

\begin{figure}[p]
\centering\includegraphics[width=\textwidth]{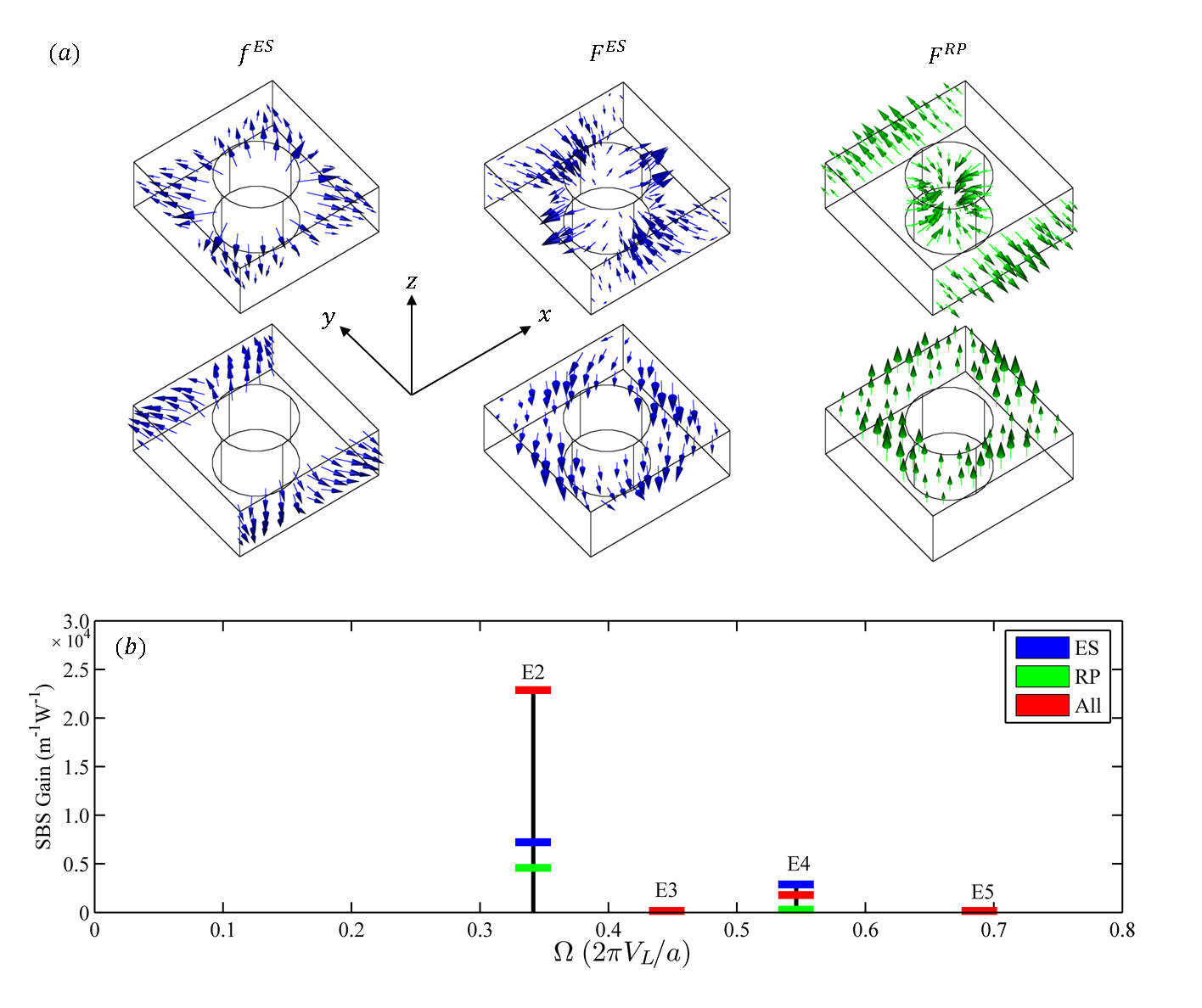}
\caption{Optical force distributions and gain coefficients of forward SBS. The operating point is $\omega = 0.265(2\pi c/a)$ and $k = 0.75(\pi / a)$ with $a=420$nm. Elastic modes at $q=0$ are excited. (a) Distributions of optical forces. For electrostriction body force, the two subplots show the body force density on planes $z=0$ and $y=\pm 0.4a$. For electrostriction pressure and radiation pressure, the two subplots show the pressure on the lateral surfaces and the top surface. Electrostriction pressure is multiplied by 5 so that it can be plotted on the same scale as radiation pressure. All the optical forces are symmetric about to plane $x = 0$. (b) The FSBS gains of individual elastic modes assuming $Q = 1000$. Blue, green, and red bars represent the FSBS gains under three conditions: electrostriction-only, radiation-pressure-only, and the combined effects.}
\label{fig2}
\end{figure}

In FSBS, pump and Stokes waves approximately correspond to the same optical mode $\mathbf{E}e^{-i\omega t}$, and excite standing-wave elastic modes at $q=0$. Under such conditions, electrostriction tensor in Eq. (\ref{sigma}) and MST in Eq. (\ref{T}) are given by:
\begin{eqnarray}
\sigma_{ij} &=& -\frac{1}{2}\epsilon_0 n^4 p_{ijkl} Re(E_kE_l^*) 
\label{F_sigma} \\
T_{ij} &=& \frac{1}{2}\epsilon_0 \epsilon
(2Re(E_iE_j^*)-\delta_{ij}|\mathbf{E}|^2)
\label{F_T} 
\end{eqnarray}
Both electrostriction tensor and MST are real-values, resulting in real-valued optical forces. We select an operating point at $\omega = 0.265(2\pi c/a)$ and $k = 0.75(\pi / a)$ with $a=420$nm, and compute the distributions of electrostriction body force, electrostriction pressure, and radiation pressure (Fig. \ref{fig2}(a)). The dominant component of electrostriction body force is $f^{ES}_y$, because the dominant component of electric field is $E_y$, and $p_{11}$ is about five times larger than $p_{12}$. Radiation pressure point outwards, which is about five times larger than electrostriction pressure.

One important feature about the optical force distribution is that all optical forces are symmetric about plane $x=0$ although the optical eigen-mode $\mathbf{E}$ doesn't have this symmetry. This can be formerly proven by examining the symmetry of electrostriction tensor and MST. Because the crystal structure of the waveguide material is symmetric about $x=0$, the photo-elastic tensor is zero when there is odd number of $x$ in the subscript: $p_{ijkl} = p_{ijkl}s_i s_j s_k s_l$. Using this property and $s_i^2=1$, we have
\begin{eqnarray}
\sigma_{ij}(-x,y,z) 
&=& - \frac{1}{2}\epsilon_0 n^4 p_{ijkl} Re(E_{k}(-x,y,z)E_{l}^*(-x,y,z)) \nonumber \\
&=& - \frac{1}{2}\epsilon_0 n^4 p_{ijkl} Re(E_{k}^*(x,y,z)E_{l}(x,y,z)) s_i s_j s_k s_l s_k s_l \nonumber \\
&=& \sigma_{ij}(x,y,z)s_is_j 
\label{PF_sigma}
\end{eqnarray}
Similarly, using $\delta_{ij}s_is_j=\delta_{ij}$, we have
\begin{eqnarray}
T_{ij}(-x,y,z) 
&=& \frac{1}{2}\epsilon_0 \epsilon (2Re(E_i(-x,y,z)E_j^*(-x,y,z))-\delta_{ij}|\mathbf{E}(-x,y,z)|^2) \nonumber \\
&=& \frac{1}{2}\epsilon_0 \epsilon
(2Re(E_i(x,y,z)E_j^*(x,y,z))-\delta_{ij}|\mathbf{E}(x,y,z)|^2)s_is_j \nonumber \\
&=& T_{ij}(x,y,z)s_is_j
\label{PF_T}
\end{eqnarray}
Combining Eq. (\ref{PF_sigma}) and (\ref{PF_T}) with the fact that optical force is given by the divergence of the corresponding tensor, we conclude that both optical forces in FSBS are symmetric about plane $x=0$. 

The symmetry of optical forces, together with symmetry property of elastic modes at $q=0$, indicates that only xeven modes are excitable. We calculate the FSBS gain assuming a mechanical quality factor $Q=1000$ for all the elastic modes. As expected, only xeven modes E2 and E4 have nonzero FSBS gains (Fig. \ref{fig2}(b)). Mode E2 has large displacement in $y$ direction and small displacement in $z$ direction. Such modal profile agrees well with electrostriction body force and the radiation pressure on the outer lateral surfaces, generating large FSBS gains from electrostriction ($0.71 \times 10^4$m\textsuperscript{-1}W\textsuperscript{-1}) and radiation pressure ($0.44 \times 10^4$m\textsuperscript{-1}W\textsuperscript{-1}). Furthermore, these two effect add up constructively, resulting a total FSBS gain as large as $2.27 \times 10^4$m\textsuperscript{-1}W\textsuperscript{-1}. Mode E4 has a small FSBS gain of $0.17 \times 10^4$m\textsuperscript{-1}W\textsuperscript{-1} because (1) the nodal planes of $u_y$ reduce the overlap integral of electrostriction body force and (2) radiation pressures on the outer and inner lateral surfaces are canceled out to a large extent.

\begin{figure}[p]
\centering\includegraphics[width=\textwidth]{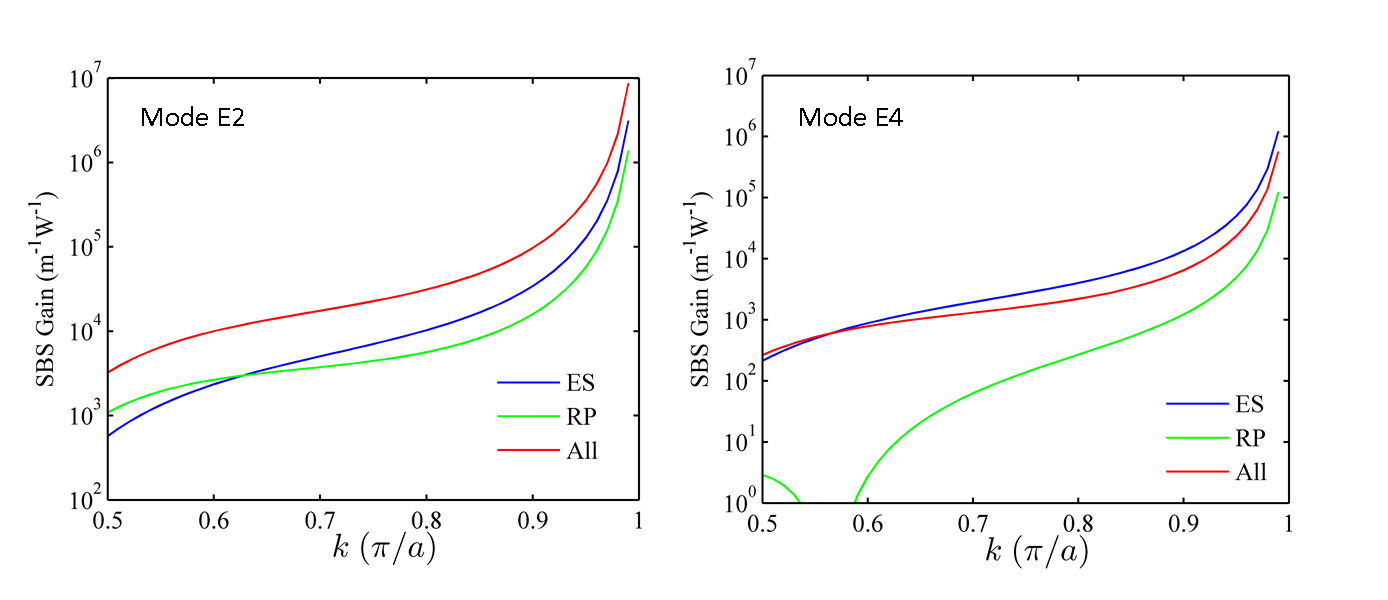}
\caption{Forward SBS gains of mode E2 and mode E4 as $k$ varies from $0.5 (\pi / a)$ to $\pi / a$, and the corresponding $a$ varies from 354nm to 434nm. For both mode E2 and E4, the FSBS gains approach to infinity as $O(1/\Delta k^2)$ at the brillouin zone boundary.}
\label{fig3}
\end{figure}

Next, we study how the FSBS gain varies as the operating point moves from brillouin zone interior to boundary (Fig. \ref{fig3}). For mode E2, electrostriction force and radiation pressure always add up constructively, creating an even larger FSBS gain. For mode E4, the radiation-pressure-only gain coefficient vanishes around $k=0.57(\pi/a)$, because of the cancellation of radiation pressures on different surfaces. For both mode E2 and E4, when the operating point approaches brillouin zone boundary, the overlap integrals approach to constants while the optical group velocity vanishes as $O(\Delta k)$ ($\Delta k = |k - \pi/a|$). As a result, the FSBS gains approach to infinity as $O(1/ \Delta k ^2)$.

\section{Backward SBS}

\begin{figure}[p]
\centering\includegraphics[width=\textwidth]{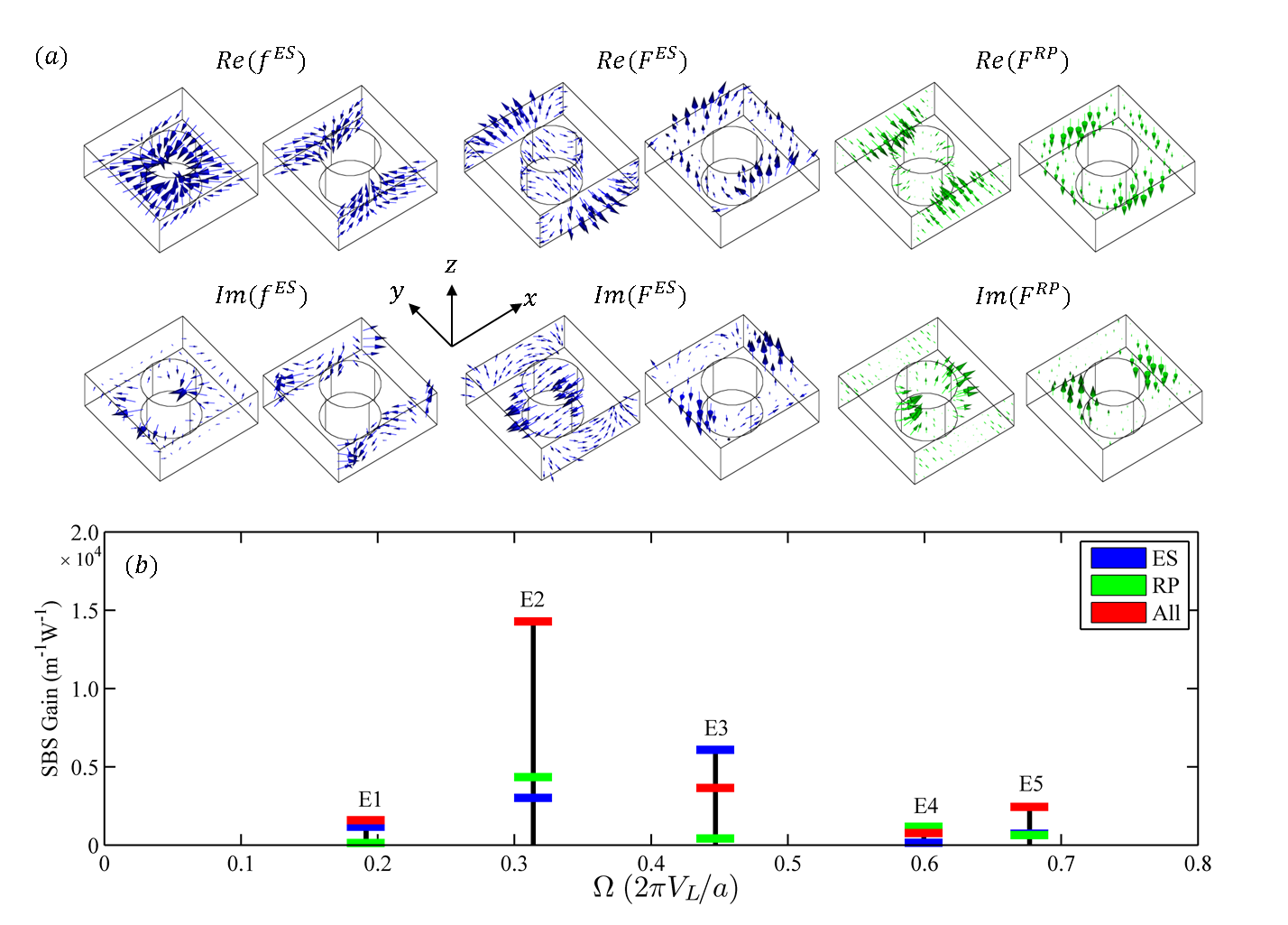}
\caption{Optical force distributions and gain coefficients of backward SBS. The operating point is $\omega = 0.265(2\pi c/a)$ and $k = 0.75(\pi / a)$ with $a=420$nm. Elastic modes at $q = 1.5 (\pi / a)$ are excited. (a) Distributions of the real and imaginary parts of optical forces. For electrostriction body force, the two subplots show the body force density on planes $z=0$ and $y=\pm 0.4a$. For electrostriction pressure and radiation pressure, the two subplots show the pressure on the lateral surfaces and the top surface. Electrostriction pressure is multiplied by 5 so that it can be plotted in the same scale as radiation pressure. The real part of optical forces is symmetric about plane $x = 0$, while the imaginary part of optical forces is anti-symmetric about plane $x=0$. (b) The BSBS gains of individual elastic modes assuming $Q=1000$. Blue, green, and red bars represent the BSBS gains under three conditions: electrostriction-only, radiation-pressure-only, and the combined effects.}
\label{fig4}
\end{figure}

In BSBS, pump and Stokes waves travel in the opposite directions, exciting elastic modes at $q=2k$. Under such conditions, electrostriction tensor in Eq. (\ref{sigma}) and MST in Eq. (\ref{T}) are given by:
\begin{eqnarray}
\sigma_{ij} &=& -\frac{1}{2}\epsilon_0 n^4 p_{ijkl} E_kE_l 
\label{B_sigma} \\
T_{ij} &=& \frac{1}{2}\epsilon_0 \epsilon
(2E_iE_j-\delta_{ij}\mathbf{E}\cdot \mathbf{E})
\label{B_T} 
\end{eqnarray}
Both the electrostriction tensor and MST are complex-valued, resulting in complex-valued optical forces. We select an operating point at $\omega = 0.265(2 \pi c / a)$ and $k=0.75(\pi/a)$ with $a=420$nm, and compute the real and imaginary parts of electrostriction body force, electrostriction pressure, and radiation pressure (Fig. \ref{fig4}(a)). Similar to the case of FSBS, we can show that the electrostriction tensor and MST in BSBS have the following properties:
\begin{eqnarray}
\sigma_{ij}(-x,y,z) &=& \sigma_{ij}^*(x,y,z)s_is_j
\label{PB_sigma} \\
T_{ij}(-x,y,z) &=& T_{ij}^*(x,y,z)s_is_j
\label{PB_T}
\end{eqnarray}
Taking the real and imaginary parts of the expressions above, it is straightforward to show that the real (imaginary) part of optical forces is symmetric (anti-symmetric) about plane $x=0$. Under such optical forces, all the elastic modes at $q=2k$ are excitable. We calculate the BSBS gain assuming a mechanical quality factor $Q=1000$ for all the elastic modes (Fig. \ref{fig4}(b)). Mode E2 has the largest BSBS gain ($1.42 \times 10^4$m\textsuperscript{-1}W\textsuperscript{-1}), which comes from a constructive combination of electrostriction ($0.29 \times 10^4$m\textsuperscript{-1}W\textsuperscript{-1}) and radiation pressure ($0.42 \times 10^4$m\textsuperscript{-1}W\textsuperscript{-1}). 

One interesting feature about the gain coefficients is that electrostriction and radiation pressure add up either purely constructively or destructively:
\begin{equation}
G_{all} = (\sqrt{G_{ES}} \pm \sqrt{G_{RP}})^2
\end{equation}
This can also be explained using symmetry argument. As mentioned above, the real (imaginary) parts of optical forces are symmetric (anti-symmetric) about plane $x=0$. On the other hand, the real (imaginary) part of elastic displacement is even (odd) about plane $x=0$. Although both optical forces and elastic modes are complex-valued, their overlap integrals are always real. The direct interference between different optical forces can be exploited to deliberately enhance or suppress the SBS nonlinearity of certain elastic modes.

\begin{figure}[p]
\centering\includegraphics[width=\textwidth]{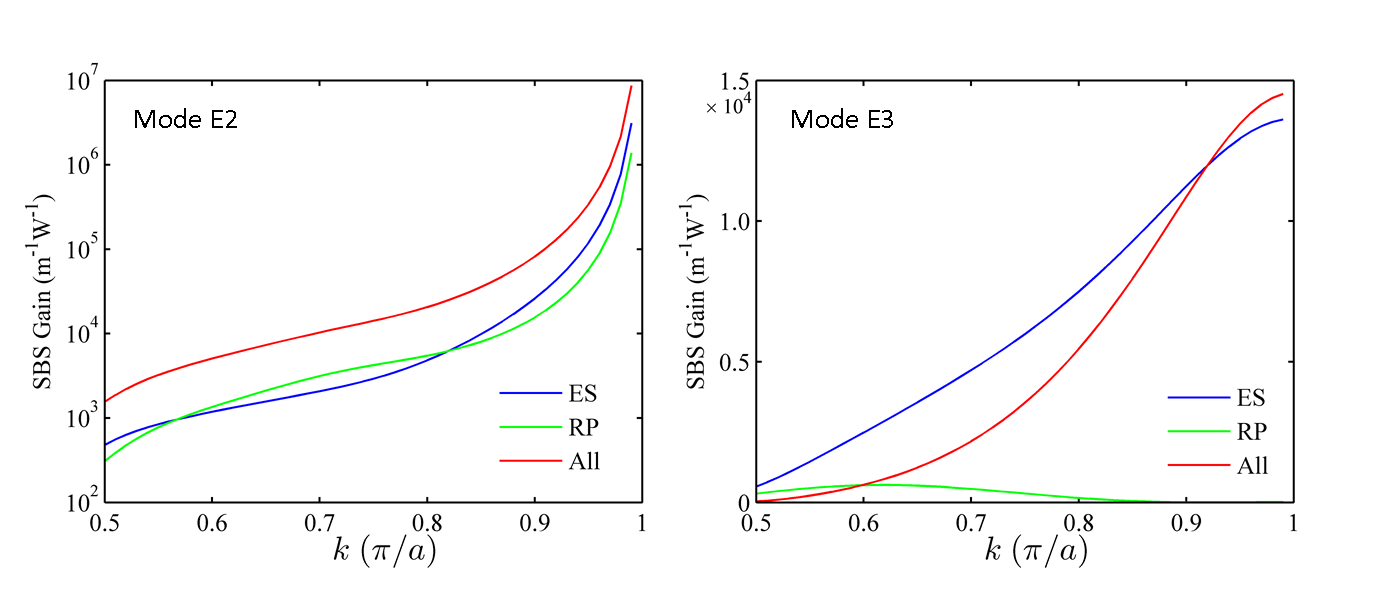}
\caption{Backward SBS gains of mode E2 and mode E3 as $k$ varies from $0.5 (\pi / a)$ to $\pi / a$ and the corresponding $a$ varies from 354nm to 434nm. For mode E2, the BSBS gain approaches to infinity as $O(1/ \Delta k^2)$. For mode E3, the BSBS gain approaches to a constant at the brillouin zone boundary.}
\label{fig5}
\end{figure}

Next, we study how the BSBS gain varies as the operating point approaches the slow light regime (Fig. \ref{fig5}). For mode E2, electrostriction force and radiation pressure always add up constructively. In the slow light regime, the BSBS gain approaches to infinity as $O(1/\Delta k ^2)$ as the optical group velocity vanishes. In contrast, the BSBS gain of mode E3 approaches to a constant at the brillouin zone boundary. This comes from two properties associated with $k=\pi/a$. At $k=\pi/a$, the optical modes at $k$ and $-k$ merge into one mode, and BSBS becomes equivalent to FSBS. So the optical forces in BSBS become symmetric about plane $x=0$. For $k$ close to $\pi / a$, the optical force in BSBS can still be decomposed into symmetric and anti-symmetric components with the anti-symmetric component on the order of $O(\Delta k)$. On the other hand, at $k=\pi/a$, $q=2k=2\pi/a$, which is equivalent to $q=0$. So the elastic modes in BSBS recover their symmetries about $x=0$. For $k$ close to $\pi / a$, elastic modes can be decomposed into odd and even components with respect to plane $x=0$. The even component in xodd modes such as mode E3 is on the order of $O(\Delta k)$. Therefore, in BSBS, the overlap integral between optical forces and mode E3 vanishes as $O(\Delta k)$. As a result, the BSBS gain approaches to a constant rather than infinity in the slow light regime.  

\section{Concluding remarks}

In this article, we analyze the forward and backward SBS processes of a periodic waveguide suspended in air. The suspended structure provides tight lateral confinement of light and nearly perfect lateral confinement of sound \cite{PRX}. The periodic structure slows down the optical group velocity. The combination of these two effects creates a giant enhancement of SBS nonlinearity over conventional nonlinear fibers \cite{Gauthier11}. In addition, we characterize elastic modes and the resultant SBS gain coefficients based their symmetries with respect to the symmetry plane perpendicular to the propagation axis. Our analysis doesn't rely on the specific waveguide geometry or the crystalline structure of silicon. The conclusion about the relation between mode symmetry and SBS gain is valid as long as (1) the waveguide has a symmetry plane perpendicular to the propagation axis and (2) the crystalline structure of the underlying material doesn't break this symmetry. Our analysis can be readily applied to simpler structures such as axially invariant waveguides and more complicated structures such as waveguide systems with photonic/phononic lateral confinement.

\newpage
\appendix
\numberwithin{equation}{section}

\section{Eigen-mode Decomposition of Elastic Response}

\subsection{Orthogonality of elastic eigen-modes}

When loss and external forces are ignored, the equation about displacement $\mathbf{u}$ is \cite{Royer1}:
\begin{equation}
\frac{\partial}{\partial x_j} c_{ijkl} \frac{\partial u_l}{\partial x_k}
= \rho\frac{\partial^2 u_i}{\partial t^2}
\end{equation}
where $\rho$ is the mass density, and $c_{ijkl}$ is the elasticity tensor. $c_{ijkl}$ is symmetric about the first two and last two indices: $c_{ijkl}=c_{jikl}$, $c_{ijkl}=c_{ijlk}$. It is also symmetric when the the first two indices and the last two indices are interchanged: $c_{klij}=c_{ijkl}$ \cite{Royer1}.

For a finite structure or an infinite structure with periodicity, the elastic eigen-modes are discrete. Considering an elastic eigen-mode with frequency $\Omega$: $\mathbf{u} = \mathbf{u} e^{-i\Omega t}$, we have:
\begin{equation}
\frac{\partial}{\partial x_j}c_{ijkl}\frac{\partial u_l}{\partial x_k} = -\Omega ^2 \rho u_i
\end{equation}
This is an eigen-equation with operator $F$ and kernel $\rho$:
\begin{equation}
F\mathbf{u} = -\Omega^2\rho \mathbf{u}
\end{equation}
We define the inner product between two vector fields as the overlap integral over the finite structure or the unit cell of an infinite structure with periodicity:
\begin{equation}
\langle \mathbf{A},\mathbf{B} \rangle = \int A_i^* B_i dV
\end{equation}
By integrating by parts and using the properties of $c_{ijkl}$, we can show that operator $F$ is Hermitian:
\begin{eqnarray}
\langle \mathbf{A},F\mathbf{B} \rangle 
&=& \int A_i^* \frac{\partial}{\partial x_j}c_{ijkl}\frac{\partial B_l}{\partial x_k} dV \nonumber \\
&=& \int c_{ijkl} \frac{\partial A_i^*}{\partial x_j} \frac{\partial B_l}{\partial x_k} dV \nonumber \\
&=& \int c_{ijkl} \frac{\partial B_i}{\partial x_j} \frac{\partial A_l^*}{\partial x_k} dV \nonumber \\
&=& \int B_i \frac{\partial}{\partial x_j}c_{ijkl}\frac{\partial A_l^*}{\partial x_k} dV \nonumber \\
&=& \langle F\mathbf{A},\mathbf{B} \rangle
\end{eqnarray}
By the property of Hermitian operators, elastic eigen-modes are orthogonal under kernel $\rho$:
\begin{equation}
\langle \mathbf{u}_m, \rho \mathbf{u}_n \rangle = \delta_{mn}
\langle \mathbf{u}_m, \rho \mathbf{u}_m \rangle
\end{equation}

\subsection{Elastic response with loss and external forces}

Now we apply external force $\mathbf{f}$ to the structure. The elastic equation becomes:
\begin{equation}
\frac{\partial}{\partial x_j} c_{ijkl} \frac{\partial u_l}{\partial x_k} + f_i = \rho\frac{\partial^2 u_i}{\partial t^2}
\end{equation}
Assuming $f$ is time-harmonic with frequency $\Omega$, we have
\begin{equation}
\frac{\partial}{\partial x_j} c_{ijkl} \frac{\partial u_l}{\partial x_k} + f_i = -\rho \Omega^2 u_i
\end{equation}
By decomposing $\mathbf{u}$ into the eigen-modes $\mathbf{u}=\sum_m b_m \mathbf{u}_m$, we get:
\begin{equation}
\sum_m b_m \rho \left( \Omega_m^2 - \Omega^2 \right) \mathbf{u}_m = \mathbf{f}
\end{equation}
Taking the inner product of $\mathbf{u}_m$ with both sides of the equation above and applying the orthogonality condition, we have:
\begin{equation}
b_m = \frac{\langle \mathbf{u}_m, \mathbf{f} \rangle}
{\langle \mathbf{u}_m,\rho \mathbf{u}_m \rangle}
\frac{1}{\Omega_m^2 - \Omega^2}
\end{equation}

We now consider a more general case, where the mechanical loss is present. The commonly encountered mechanical loss mechanism includes air damping, thermoelastic dissipation, and clamping losses \cite{Kenny08}. The first order effect of loss can be captured by adding an imaginary part to $\Omega_m$: $\Omega_m - i \Gamma_m/2$. The mechanical quality factor is defined as $Q_m = \Omega_m / \Gamma_m$. Assuming $Q_m$ is well above 1, we have,
\begin{equation}
b_m = \frac{\langle \mathbf{u}_m, \mathbf{f} \rangle}
{\langle \mathbf{u}_m,\rho \mathbf{u}_m \rangle}
\frac{1}{\Omega_m \Gamma_m}
\frac{\Gamma_m/2}{\Omega_m - \Omega - i\Gamma_m/2}
\label{b_m}
\end{equation}

\section{Calculation of Optical Forces}

\subsection{Electrostriction forces}

Electrostriction force is derived from electrostriction tensor. The instantaneous electrostriction tensor is given by:
\begin{equation}
\sigma_{ij} = - \frac{1}{2}\epsilon_0 n^4 p_{ijkl} E_k E_l
\label{sigma}
\end{equation}
where $n$ is the refractive index, and $p_{ijkl}$ is the photoelastic tensor \cite{Royer2}. When both pump and Stokes waves are present, the total electric field is given by $(\mathbf{E}_p e^{-i\omega_p t} + \mathbf{E}_s e^{-i\omega_s t})/2 + c.c$. Inserting this expression to Eq. (\ref{sigma}), and taking out the components with frequency $\Omega$, we get the time-harmonic electrostriction tensor $\sigma_{ij} e^{-i\Omega t}$:
\begin{equation}
\sigma_{ij} = - \frac{1}{4}\epsilon_0 n^4 p_{ijkl} (E_{pk}E_{sl}^* + E_{pl}E_{sk}^*)
\end{equation}

Electrostriction force is the divergence of electrostriction tensor. In a system consisting of domains of homogeneous materials, electrostriction forces can exist within each material (electrostriction body force) and on the interfaces (electrostriction pressure). Electrostriction body force is given by:
\begin{equation}
f_i^{ES} = - \partial_j \sigma_{ij}
\label{f}
\end{equation}
Electrostriction pressure on the interface between material 1 and 2 is given by (normal vector $n$ points from 1 to 2):
\begin{equation}
F_i^{ES} = (\sigma_{1ij} - \sigma_{2ij}) n_j
\label{FES}
\end{equation}
Equation (\ref{FES}) shows that electrostriction pressure can have tangent components on material boundaries.

\subsection{Radiation pressure}

Radiation pressure is derived from Maxwell Stress Tensor (MST). For a dielectric system ($\mu=1$) without free charges ($\rho=0,J=0$), radiation pressure is localized where the gradient of $\epsilon$ is nonzero \cite{Gordan73,Johnson02}. For a system consisting of homogeneous materials, radiation pressure only exists on the interfaces. The electric part of instantaneous MST is:
\begin{equation}
T_{ij} = \epsilon_0 \epsilon (E_i E_j - \frac{1}{2}\delta_{ij}E^2)
\label{T}
\end{equation}
When both pump and Stokes waves are present, the total electric field is given by $(\mathbf{E}_p e^{-i\omega_p t} + \mathbf{E}_s e^{-i\omega_s t})/2 + c.c$. Inserting this expression into Eq. (\ref{T}) and filtering out the component with frequency $\Omega$, we arrive at the time-harmonic MST $T_{ij}e^{-i\Omega t}$:
\begin{equation}
T_{ij} = \frac{1}{2}\epsilon_0 \epsilon (E_{pi}E_{sj}^*+E_{pj}E_{si}^*-\delta_{ij}\mathbf{E}_p \cdot \mathbf{E}_s^*)
\end{equation}

Radiation pressure on the interface between material 1 and 2 is given by the discontinuity of MST:
\begin{equation}
F_i^{RP} = (T_{2ij}-T_{1ij})n_j
\end{equation}
By decomposing the electric field into its normal and tangent components with respect to the dielectric interfaces $\mathbf{E} = E_n\mathbf{n} + E_t\mathbf{t}$, and using the boundary condition $\epsilon_1 E_{1n} = \epsilon_2 E_{2n} = D_n$ and $E_{1t} = E_{2t} = E_t$, we can get the time-harmonic radiation pressure $\mathbf{F}^{RP} e^{-i \Omega t)}$:
\begin{equation}
\mathbf{F}^{RP} = -\frac{1}{2}\epsilon_0 E_{pt}E_{st}^* (\epsilon_2 - \epsilon_1) \mathbf{n}
+ \frac{1}{2}\epsilon_0^{-1} D_{pn}D_{sn}^* (\epsilon_2^{-1} - \epsilon_1^{-1}) \mathbf{n}
\label{FRP}
\end{equation}
Equation (\ref{FRP}) shows that radiation pressure is always normal to the interface.

\newpage

\end{document}